\documentclass[traditabstract]{aa}

\usepackage{graphicx}
\usepackage{txfonts}
\begin{document}
\renewcommand{\textfraction}{0.0}
\renewcommand{\bottomfraction}{1.0}
\renewcommand{\dblfloatpagefraction}{1.00}
\title{Submillimeter continuum observations of Sagittarius B2 at subarcsecond
spatial resolution}

\author{
S.-L. Qin,\inst{1} P. Schilke,\inst{1} R. Rolffs,\inst{1,2} C. Comito,\inst{2}
D.C. Lis\inst{3} and Q. Zhang\inst{4}
}

\institute{I. Physikalisches Institut, Universit\"at zu K\"oln,
             Z\"ulpicher Str. 77, 50937 K\"oln, Germany \\
            \email{qin@ph1.uni-koeln.de, schilke@ph1.uni-koeln.de}
\and  Max-Planck-Institut f\"ur Radioastronomie, Auf dem H\"ugel 69, 53121 Bonn,
Germany
\and California Institute of Technology, Cahill Center for Astronomy and
Astrophysics 301-17, Pasadena, CA 91125 USA
\and Harvard-Smithsonian Center for Astrophysics, 60 Garden Street, Cambridge MA
02138, USA
}

\abstract{We report the first high spatial resolution submillimeter continuum
observations of the Sagittarius~B2 cloud complex using the Submillimeter Array
(SMA). With the subarcsecond resolution provided by the SMA, the two massive
star-forming clumps Sgr~B2(N) and Sgr~B2(M) are resolved into multiple compact sources. In total, twelve submillimeter cores are identified in the Sgr~B2(M)
region, while only two components are observed in the Sgr~B2(N) clump. The gas
mass and column density are estimated from the dust continuum emission. We find
that most of the cores have gas masses in excess of 100 M$_{\odot}$ and column
densities above 10$^{25}$~cm$^{-2}$. The very fragmented appearance of
Sgr~B2(M), in contrast to the monolithic structure of Sgr B2 (N), suggests that
the former is more evolved. The density profile of the Sgr~B2(N)-SMA1 core is well
fitted by a Plummer density distribution. This would lead one to believe that
in the evolutionary sequence of the Sgr B2 cloud complex, a massive star forms first in an homogeneous core,
and the rest of the cluster forms subsequently in the then fragmenting
structure.}

  \titlerunning{Submillimeter continuum observations of Sgr~B2}
  \authorrunning{S.-L. Qin et al.}
  \maketitle
%
\section{Introduction}
The Sagittarius B2 star-forming region is located $\sim$ 100 pc from Sgr A$^*$,
within the $\sim 400$ pc wide dense central molecular zone (CMZ) of the Galactic
center, at a distance of $\sim$8 kpc from the Sun (Reid et al. 2009). It is the
strongest submillimeter continuum source in the CMZ (Schuller et al. 2009). It
contains dense cores, Sgr~B2(N) and Sgr~B2(M), hosting clusters of compact H{\sc
ii} regions (Gaume et al. 1995; de Pree, Goss \& Gaume 1998). It has been suggested that these two hot cores are at different evolutionary
stages (Reid et al 2009; Lis et al. 1993; Hollis et al. 2003; Qin et al. 2008).
Spectral observations in centimeter and millimeter regimes have been conducted
towards Sgr~B2 ({\it e.g.} Carlstrom \& Vogel 1989; Mehringer \& Menten 1997; Nummelin et al. 1998; Liu \& Snyder 1999; Hollis et al. 2003; Friedel et al. 2004; Jones et al. 2008; Belloche et al. 2008), suggesting that Sgr~B2(N) is
chemically more active. Nearly half of all known interstellar molecules were
first identified in Sgr~B2(N), although sulphur-bearing molecules are more
abundant in Sgr~B2(M) than in Sgr~B2(N). 
                                  
The differences between Sgr~B2(N) and Sgr~B2(M), in terms of both kinematics and
chemistry, may originate from different physical conditions and thus different
chemical histories, or may simply be an evolutionary effect. A clearer
understanding of the small-scale source structure and the exact origin of the
molecular line emission is needed to distinguish between these two
possibilities.  In this Letter, we present high spatial resolution submillimeter
continuum observations of Sgr~B2(N) and Sgr~B2(M), using the SMA\footnote {The
Submillimeter Array is a joint project between the Smithsonian Astrophysical
Observatory and the Academia Sinica Institute of Astronomy and Astrophysics, and
is funded by the Smithsonian Institution and the Academia Sinica.}. The
observations presented here resolve both of the Sgr~B2 clumps into
multiple submillimeter components. Combining with the SMA spectral line 
data
cubes and ongoing Herschel/HIFI complete spectral surveys towards Sgr~B2(N) and
Sgr~B2(M) in the HEXOS key project, these observations will help us to answer fundamental
questions about the chemical composition and physical conditions in Sgr~B2(N)
and Sgr~B2(M).

\section{Observations}
The SMA observations of Sgr~B2 presented here were carried out using seven
antennas in the compact configuration on 2010 June 11, and using eight antennas
in the very extended configuration on 2010 July 11. The phase tracking centers
were $\alpha\,( \mathrm{J}\,2000.0)
=17^{\mathrm{h}\,}47^{\mathrm{m}\,}19.883^{\mathrm{s}\,} , \delta\,(
\mathrm{J}\,2000.0) =-28^{\circ }22^{\prime }18.4^{\prime \prime }$ for
Sgr~B2(N) and $\alpha\,( \mathrm{J}\,2000.0)
=17^{\mathrm{h}\,}47^{\mathrm{m}\,}20.158^{\mathrm{s}\,} , \delta\,(
\mathrm{J}\,2000.0) =-28^{\circ }23^{\prime }05.0^{\prime \prime }$ for
Sgr~B2(M). Both tracks were observed in double-bandwidth mode with a 4~GHz bandwidth
in each of the lower sideband (LSB) and upper sideband (USB). The spectral
resolution was 0.8125~MHz per channel, corresponding to a velocity resolution of
$\sim$0.7 km~s$^{-1}$.  The observations covered rest frequencies from 342.2 to
346.2~GHz (LSB), and from 354.2 to 358.2 GHz (USB). Observations of QSOs 1733-130 and 1924-292
were evenly interleaved with the array pointings toward Sgr~B2(N) and Sgr~B2(M)
during the observations in both configurations, to perform antenna gain calibration.

For the compact configuration observations, the typical system temperature was
273 K. Mars, QSOs 3c454.3, and 3c279 were observed for bandpass calibration. The flux
calibration was based on the observations of Neptune ($\sim$1.1$^{\prime \prime
}$). For the very extended array observations, the typical system temperature was 292~K. Both 3c454.3 and 3c279 were used for
bandpass calibration, and Uranus ($\sim$1.7$^{\prime \prime }$) was used for
flux calibration. The absolute flux scale is estimated to be accurate to within
20\%.

The calibration and imaging were performed in Miriad (Sault, Teuben \& Wright
1995). We note that there are spectral-window-based bandpass errors in both amplitude and
phase on some baselines in the compact array data, which were corrected by
use of a bright point source using the BLCAL task. The system temperature measurements for
antennas 2 and 7 in the very extended array data were not recorded properly and
were corrected using the SMAFIX task. As done by Qin et al. (2008), we selected line free channels which may contain some contributions from weak and densely spaced blended lines, although at least obvious line contributions were excluded. The continuum images were constructed from those, combining the LSB and USB data of both the compact and very extended
array observations. We performed a self-calibration on the continuum data using
the model from `CLEANed' components for a few iterations to remove
residual errors. Using the model from `CLEANed' components, the self-calibration
in our case did not introduce errors in the source structure, but improved the
image quality and minimized the gain calibration errors. The final images were
corrected for the primary beam response. The projected baselines ranged from 9
to 80 k$\lambda$ in the compact configuration and from 27 to 590 k$\lambda$ in
the very extended configuration. The resulting synthesized beam is
0$\rlap{.}^{\prime\prime}$4$\times$0$\rlap{.}^{\prime\prime}$24
(PA=14.4$^{\circ}$) using uniform weighting, and the 1$\sigma$ rms noise levels
are 21 and 31 mJy for Sgr~B2(M) and Sgr~B2(N) images, respectively. The
difference in rms noise is caused by having more line free
channels for continuum images in Sgr~B2(M) (2417 channels) than in Sgr~B2(N) (740 channels). Since there are no systematic offsets between the submillimeter and cm sources, we believe that the absolute astrometry is as good as 0.1$^{\prime\prime}$.

\section{Results}
Continuum images of Sgr~B2(N) and Sgr~B2(M) at 850 $\mu$m are shown in Figure~1.
Multiple submillimeter continuum cores are clearly detected and resolved
towards both Sgr~B2(M) and Sgr~B2(N) with a spatial resolution of
0$\rlap{.}^{\prime\prime}$4$\times$0$\rlap{.}^{\prime\prime}$24. Unlike in
radio observations at 1.3 cm with a comparable resolution, which detected the UC~H{\sc
ii} regions K1, K2, K3, and K4 (Gaume et al. 1995), only two
submillimeter continuum sources, SMA1 and SMA2, are observed in
Sgr~B2(N) (Fig.~1). The bright component
Sgr~B2(N)-SMA1 is situated close
to the UC~H{\sc ii} region K2, and Sgr~B2(N)-SMA2 is located $\sim$5$^{\prime\prime}$ north of Sgr~B2(N)-SMA1. The
observations showed that large saturated molecules exist only within a small
region ($< 5^{\prime\prime}$) of Sgr~B2(N) called the Large Molecule Heimat, Sgr~B2(N-LMH) (Snyder, Kuan \& Miao 1994). Our current observations indicate
that Sgr~B2(N-LMH) coincides with Sgr~B2(N)-SMA1. Sgr~B2(N)-SMA2 was
also detected in continuum emission at 7~mm and 3~mm (Rolffs et al. 2011; Liu \& Snyder 1999) and molecular lines of CH$_3$OH and C$_2$H$_5$CN at 7 mm
(Mehringer \& Menten 1997; Hollis et al. 2003). Lower resolution continuum observations at 1.3 mm (see note to Table 1 of Qin et al. 2008) suggested that the Sgr B2(N)K1-K3 clump could not be fitted with a single Gaussian component, and another component existed at Sgr~B2(N)-SMA2 position. Our observations here resolved out Sgr~B2(N)-SMA2, confirming it to be a high-mass core.

\begin{figure*}[!t] \centering \includegraphics[width=7.81cm,angle=-90]{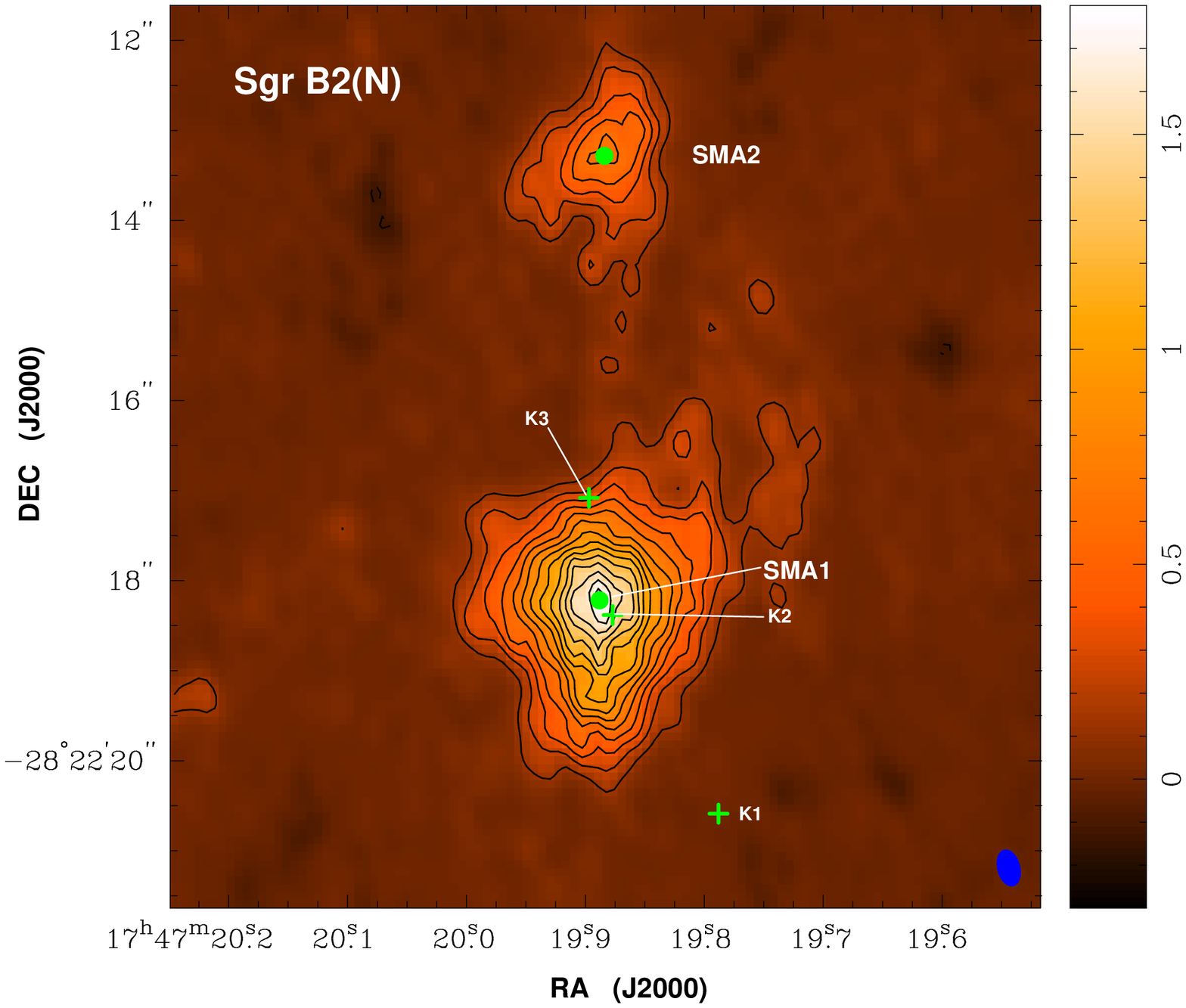}
\includegraphics[width=7.81cm,angle=-90]{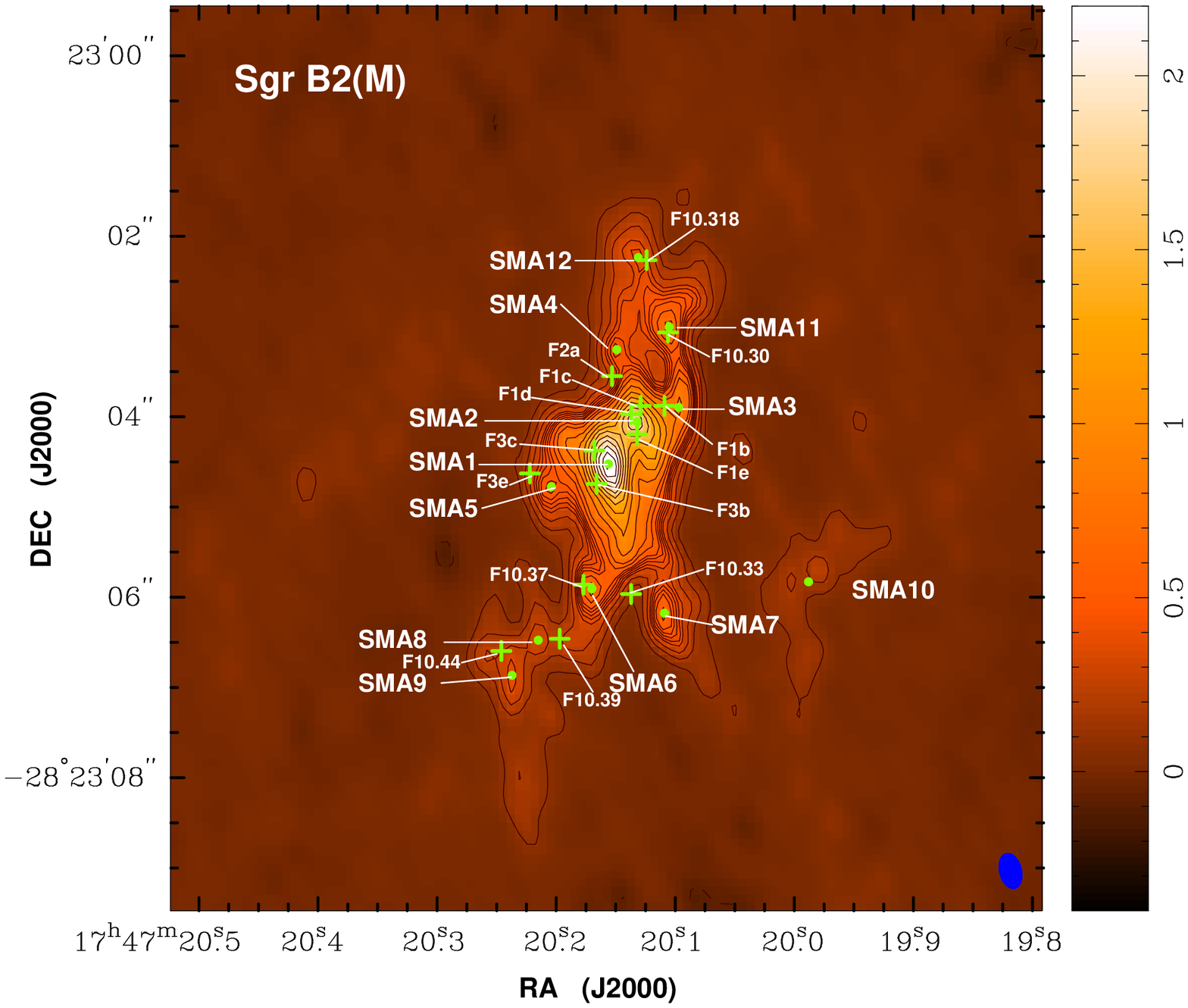} \caption{Continuum maps of
Sgr~B2 at 850~$\mu$m, with a synthesized beam of
0$\rlap{.}^{\prime\prime}$4$\times$0$\rlap{.}^{\prime\prime}$24,
PA=14.4$^{\circ}$ (lower-right corner in each panel). The left panels shows the
image of Sgr~B2(N), with contour levels ($-$1, 1, \ldots 14)$\times$4$\sigma$ (1$\sigma$=0.031 Jy~beam$^{-1}$). The cross
symbols indicate the positions of UC~H{\sc ii} regions detected in 1.3~cm
continuum (Gaume et al. 1995). The right panel present the image of Sgr~B2(M),
with contour levels ($-$1, 1, 2, 3, 4, 4.5, 5.5, 5, 6, 7, 8, 10, \ldots 28)$\times$4$\sigma$ (1$\sigma$=0.021 Jy~beam$^{-1}$). The cross
symbols indicate the positions of UC~H{\sc ii} regions detected in 7 mm
continuum (de Pree, Goss \& Gaume 1998). In each panel, the filled circle
symbols present the peak positions of the submillimeter continuum sources.
}\label{}
\end{figure*}

The continuum image of Sgr~B2(M) (Fig. 1, right panel) shows a complicated
morphology, with a roughly north-south extending envelope encompassing several
compact components. In total, twelve submillimeter sources are resolved in
Sgr~B2(M). Using the Very Large Array (VLA), Gaume et al. (1995) detected
four bright UC~H{\sc ii} regions (F1--F4) at 1.3 cm within 2$^{\prime\prime}$ in
Sgr~B2(M). The highest resolution 7~mm VLA image with a position accuracy of $0.1^{\prime\prime}$ (de Pree, Goss \& Gaume 1998) resolved nineteen UC~H{\sc ii} regions in the central region of Sgr~B2(M)(F1--F4). Five submillimeter components, Sgr~B2(M)-SMA1 to SMA5, are detected in the central region of Sgr~B2(M) in our observations. Outside the central region, seven components, Sgr B2(M)-SMA6 to SMA12 are identified.  Given the positional accuracies of our observations and the observations by de Pree, Goss \& Gaume (1998), the projected positions of Sgr B2(M)-SMA2, SMA6, SMA11 and SMA12 coincide with those of the UC H{\sc II} regions F1, F10.37, F10.30, and F10.318, respectively. No centimeter source is detected towards Sgr~B2(M)-SMA10, which is located south-west of the central region and displays an extended structure.

\begin{table*}[t]
\caption{Properties of the Continuum Sources}
\small
\centering
\begin{tabular}{cccccccc}
\hline\hline\noalign{\smallskip}
Source & $\alpha$ (J2000.0) & $\delta$ (J2000.0) & Deconvolved size& Peak
Intensity & Flux Density&  $M_{\rm H_{2}}$ & $N_{\rm H_{2}}$  \\
& & &&~~~(Jy~beam$^{-1}$)&(Jy)&  (10$^{2}$M$_{\odot}$)& ($10^{25}$cm$^{-2}$) \\
\noalign{\smallskip}\hline

Sgr~B2(N)-SMA1 & 17 47 19.889  &--28 22
18.22&1$\rlap{.}^{\prime\prime}72\times1\rlap{.}^{\prime\prime}28
(-7.7^{\circ}$)&1.79$\pm$0.039&47.48$\pm$1.034&27.31$\pm$0.59&4.54$\pm$0.1 \\

Sgr~B2(N)-SMA2 & 17 47 19.885  &--28 22
13.29&1$\rlap{.}^{\prime\prime}44\times1\rlap{.}^{\prime\prime}02
(-24.5^{\circ}$) & 0.601$\pm$0.031&10.12$\pm$0.521&5.82$\pm$ 0.38&1.45$\pm$0.1
\\
Sgr~B2(M)-SMA1 & 17 47 20.157  &--28 23
04.53&1$\rlap{.}^{\prime\prime}51\times0\rlap{.}^{\prime\prime}59
(10.6^{\circ}$)&2.39$\pm$0.138&20.9$\pm$1.21&12.02$\pm$0.7 &4.94$\pm$0.29   \\
Sgr~B2(M)-SMA2 & 17 47 20.133  &--28 23
04.06&0$\rlap{.}^{\prime\prime}98\times0\rlap{.}^{\prime\prime}58
(14.7^{\circ}$) & 1.75$\pm$0.12&12.51$\pm$0.859&7.19$\pm$0.49 &4.63$\pm$0.32  \\
Sgr~B2(M)-SMA3 & 17 47 20.098  &--28 23
03.9&0$\rlap{.}^{\prime\prime}93\times0\rlap{.}^{\prime\prime}3 (-9.9^{\circ}$)
& 0.866$\pm$0.09&4.147$\pm$0.431&2.38$\pm$0.25 &3.13$\pm$0.33   \\
Sgr~B2(M)-SMA4 & 17 47 20.150  &--28 23
03.26&0$\rlap{.}^{\prime\prime}55\times0\rlap{.}^{\prime\prime}32
(-1.5^{\circ}$)&0.483$\pm$0.021&1.443$\pm$0.063&0.83$\pm$0.04 &1.73$\pm$0.08  
\\
Sgr~B2(M)-SMA5 & 17 47 20.205  &--28 23
04.78&0$\rlap{.}^{\prime\prime}66\times0\rlap{.}^{\prime\prime}52
(15.9^{\circ}$) & 0.52$\pm$0.051&2.488$\pm$0.244&1.43$\pm$0.14 &1.53$\pm$0.15  
\\
Sgr~B2(M)-SMA6 & 17 47 20.171  &--28 23
05.91&0$\rlap{.}^{\prime\prime}75\times0\rlap{.}^{\prime\prime}43
(-24.3^{\circ}$) & 0.71$\pm$0.067&3.05$\pm$0.288&1.75$\pm$0.17 &1.99$\pm$0.19  
\\
Sgr~B2(M)-SMA7 & 17 47 20.11  &--28 23
06.18&0$\rlap{.}^{\prime\prime}72\times0\rlap{.}^{\prime\prime}37
(14.9^{\circ}$)&0.6$\pm$0.051&2.366$\pm$0.201&1.36$\pm$0.12 &1.87$\pm$0.16  \\
Sgr~B2(M)-SMA8 & 17 47 20.216  &--28 23
06.48&0$\rlap{.}^{\prime\prime}52\times0\rlap{.}^{\prime\prime}45
(-63.4^{\circ}$) & 0.31$\pm$0.048&1.246$\pm$0.193&0.71$\pm$0.11 &1.12$\pm$0.17  
\\
Sgr~B2(M)-SMA9 & 17 47 20.238  &--28 23
06.87&0$\rlap{.}^{\prime\prime}83\times0\rlap{.}^{\prime\prime}42
(23.9^{\circ}$)&0.369$\pm$0.032&1.922$\pm$0.164&1.11$\pm$0.09 &1.16$\pm$0.1   \\
Sgr~B2(M)-SMA10 & 17 47 19.989  &--28 23
05.83&3$\rlap{.}^{\prime\prime}06\times1\rlap{.}^{\prime\prime}0
(-25.8^{\circ}$)&0.183$\pm$0.015&4.321$\pm$0.363&2.49$\pm$0.29 &0.3$\pm$0.03  
\\
Sgr~B2(M)-SMA11 & 17 47 20.106  &--28 23
03.01&0$\rlap{.}^{\prime\prime}97\times0\rlap{.}^{\prime\prime}48
(-27.7^{\circ}$)&0.695$\pm$0.058&4.423$\pm$0.37&2.54$\pm$0.21 &2$\pm$0.17     \\
Sgr~B2(M)-SMA12 & 17 47 20.132  &--28 23
02.24&0$\rlap{.}^{\prime\prime}82\times0\rlap{.}^{\prime\prime}52 (0.9^{\circ}$)
& 0.394$\pm$0.018&2.629$\pm$0.121&1.51$\pm$0.08 &1.3$\pm$0.07   \\

\noalign{\smallskip}\hline

\end{tabular}
{Note: Units of right ascension are hours, minutes, and seconds, and units of
declination are degrees, arcminutes, and arcseconds. }

\end{table*}%

 Multi-component Gaussian fits were carried out towards both the Sgr~B2(N) and
Sgr~B2 (M) clumps using the IMFIT task. The residual fluxes after fitting are 4 and 11 Jy for Sgr B2(N) and Sgr B2(M), respectively. The large residual error in Sgr B2(M) clump is most likely caused by its complicated source structure. The peak positions, deconvolved angular sizes (FWHM), peak
intensities, and total flux densities of the continuum components are summarized
in Table~1. The total flux densities of the
Sgr~B2(N) and Sgr~B2 (M) cores are $\sim$58 and 61 Jy, while the peak fluxes
measured by the bolometer array LABOCA at the 12-m telescope APEX are 150 and
138 Jy in a $18.2^{\prime\prime}$ beam, respectively (Schuller, priv. comm.),
indicating that $\sim$60\% of the flux is filtered out and that our SMA observations only
pick up the densest parts of the Sgr~B2 cores.

Assuming that the 850 $\mu$m continuum is due to optically thin dust emission
and using an average grain radius of 0.1~$\mu$m, grain density of 3~g cm$^{-3}$,
and a gas-to-dust ratio of 100 (Hildebrand 1983; Lis, Carlstrom \& Keene 1991),
the mass and column density can be calculated using the formulae given in Lis,
Carlstrom \& Keen (1991). We adopt $Q({\nu}) = 4\times$10$^{-5}$ at 850 $\mu$m
(Hildebrand 1983; Lis, Carlstrom \& Keene 1991) and a dust temperature of 150 K
(Carlstrom \& Vogel 1989; Lis et al.  1993) in the calculation. On the basis of flux densities at 1.3~cm and 7~mm (Gaume et al. 1995; de Pree, Goss \& Gaume 1998; Rolffs et al. 2011 ), most K and F subcomponents (except for K2 and F3) at 7
mm have fluxes less than or comparable  with  those  at 22.4 GHz, which produces  descending spectra and is indicative of optically thin H{\sc ii} regions at short
wavelengths. The contributions of the free-free emission to the flux densities
of the submillimeter components are smaller than 0.7\% for K2 and F3 and smaller than
0.1\% for other components, and can safely be ignored. The estimated clump masses
and column densities are given in Table 1. The flux density of the fourteen
detected submillimeter components ranges from 1.2 to 47~Jy, corresponding to gas
masses from 71 to 2731~$M_{\odot}$. The column densities are a few times
10$^{25}$ cm$^{-2}$. Under the Rayleigh-Jeans approximation, 1 Jy~beam$^{-1}$ in
our SMA observations corresponds to a brightness temperature of 113 K. The peak
brightness temperatures of Sgr~B2(M)-SMA1 and Sgr~B2(N)-SMA1 are 270 and 200 K
respectively, which place a lower limit on the dust temperatures at the peak
position of the continuum. In this paper, the column densities and masses are
estimated by use of source-averaged continuum fluxes. Based on the model fitting
(Lis et al. 1991, 1993), the adopted $Q({\nu})$, dust temperature, and optically
thin approximation are reasonable guesses for the Sgr~B2(N) and Sgr~B2(M) cloud
complexes. The total masses of Sgr~B2(N) and (M), determined by summing up over
the components, are comparable, 3313 and 3532 M$\odot$, respectively, in
spite of very different morphologies. We consider the adopted gas temperature of 150~K a reasonable guess for the average core temperatures, although some of the massive cores show higher peak brightness temperatures.  For those, the optically thin assumption is probably also not justified, and we would be underestimating their masses.

\section{Modeling}
A striking feature in the maps is the appearance of Sgr~B2(N)-SMA1, which,
although well resolved, does not appear to be fragmented. We used the
three-dimensional radiative-transfer code
RADMC-3D\footnote{http://www.ita.uni-heidelberg.de/\textasciitilde
dullemond/software/radmc-3d}, developed by C.~Dullemond, to model the continuum
emission of Sgr~B2(N)-SMA1. Figure~2 shows three example models, whose radial
profiles were obtained by Fourier-transforming the computed dust continuum maps,
`observing' with the uv coverage of the data, and averaging the image in
circular annuli. All models are heated by the stars in the UC~H{\sc ii} regions K2
and K3, assumed to have luminosities of $\approx 10^5$ L$_{\odot}$, which is uncertain, but should give the right
order of magnitude (Rolffs et al. 2011). The dust mass opacity (0.6 cm$^2$ g$^{-1}$) is interpolated from
Ossenkopf \& Henning (1994) without grain mantles or coagulation, which
corresponds to a $Q({\nu})$ of a few times 10$^{-5}$ at 850 $\mu$m and is
consistent with the value used in our calculation of masses and column densities. The model with a density distribution that follows the
Plummer profile given by $n=1.7\times 10^8 \times \left(1 + \left(\frac{r}{11500 {\rm
AU}}\right)^2\right)^{-2.5}$ H$_2$ cm$^{-3}$ (half-power radius 6500 AU)
provides the best fit. We slso show in Fig. 2 a Gaussian model with central
density $2\times 10^8$ H$_2$ cm$^{-3}$ and half-power radius 6500 AU, and a
model whose density follows a radial power law, $n=10^9\times
\left(\frac{r}{1000 {\rm AU}}\right)^{-1.5}$ H$_2$ cm$^{-3}$ (outside of 1000
AU, the radius of the H{\sc ii} region K2). The latter model reproduces the peak
flux measured by the bolometer array LABOCA at the 12-m telescope APEX, which is
150~Jy in a $18.2^{\prime\prime}$ beam, but does not fit
the inner regions observed by the SMA very well. The Plummer and Gauss models
fit Sgr~B2(N)-SMA1 well, in terms of both shape and absolute flux. This implies
that an extended component exists, which is filtered out in the SMA map but picked up by LABOCA.
\begin{figure}[h]
 \centering
 \includegraphics[width=6.8cm,angle=-90]{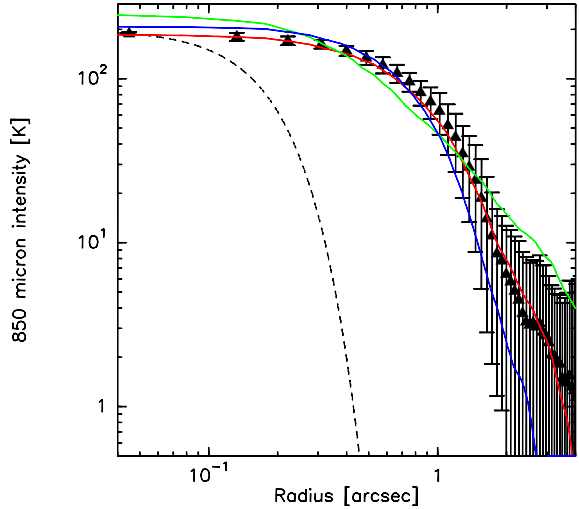} 
\caption{Radial profile of the Sgr~B2(N)-SMA1 component, with errorbars denoting
the rms in circular annuli. Overlaid are three models with different density
distributions (red: Plummer; blue: Gaussian; green: power-law). The beam is
depicted as a dashed Gaussian.}
\end{figure}

It remains unclear whether the Plummer profile, which is also used to describe density profiles of
star clusters, has any relevance to an evolving star cluster, or whether the
cluster loses its memory of the gas density profile in the subsequent dynamical
evolution. The model has a mass of around 3000
M$_\odot$ inside a radius of 11500 AU (0.056~pc), which represents an average density of $4\times 10^6$ M$_\odot$ pc$^{-3}$. This is only the
mass contained in the gas, and does not include the mass of the already formed
compact objects providing the luminosity. Sgr~B2(N) might represent a very young, embedded stage in the formation
of a massive star cluster.

\section{Discussion}
Our SMA observations resolve the Sgr~B2(M) and (N) cores into fourteen compact
 submillimeter continuum components. The two cores display very different morphologies. The
source Sgr~B2(N)-SMA1 is located north-east of the centimeter source K2, with an
offset of $\sim$0.2$^{\prime\prime}$. The continuum observations at 1.3~cm having a resolution of 0.25$^{\prime\prime}$ (Gaume et al. 1995) and at 7~mm a
resolution of 0.1$^{\prime\prime}$ (Rolffs et al. 2011) also detected a compact
component centered on K2. We failed to detect any submillimeter continuum
emission associated with sources K1, K3, and K4. Sgr B2(N)-SMA2 is a high-mass dust core.

In contrast to Sgr~B2(N), a very fragmented cluster of high mass submillimeter
sources is detected in Sgr~B2(M). In addition to the two brightest and most
massive components, Sgr~B2(M)-SMA1 and Sgr~B2(M)-SMA2, situated in the central
region of Sgr~B2(M), ten additional sources are detected, which indicates that there has been a high
degree of fragmentation. The sensitivity of 0.021 Jy~beam$^{-1}$ in our observations corresponds to a
detectable gas mass of 1.2 M$_{\odot}$, but the observations are likely to be dynamic-range-limited, so it is difficult to determine the clump mass function in
Sgr~B2(M) down to smaller masses.

The estimated column densities ($10^{25}$cm$^{-2}$=33.4~g\,cm$^{-2}$) in both
the homogeneous starforming region Sgr~B2(N) and the clustered Sgr~B2(M) region
are well in excess of the threshold of 1~g\,cm$^{-2}$ for preventing cloud
fragmentation and formation of massive stars (Krumholz \& McKee 2008). The
source sizes and masses derived from Gaussian fitting, assuming a spherical
source, lead to volume densities in excess of $10^{7}$cm$^{-3}$ for all
submillimeter sources detected in the SMA images. Assuming a gas temperature of
150~K, the thermal Jeans masses are less than 10~M$_{\odot}$, but the turbulent support is considerable.  The large column
densities, the gas masses, and the velocity field (Rolffs et al. 2010) suggest that the
submillimeter components in the two regions are gravitationally unstable and in the process of forming massive stars. This process seems more advanced in
Sgr~B2(M), which is also reflected in the large number of embedded UC~H{\sc ii} regions found there. However, star formation in Sgr~B2(N) does not appears to have progressed very far, a conclusion also supported by the presence of only one
UC~H{\sc ii} region embedded in one of the two clumps studied here. The observations also showed that massive star formation taking place in the two clumps with outflow ages of $\sim10^{3}$ and $\sim10^{4}$ years for Sgr B2(N) and Sgr B2(M) clumps, respectively (Lis et al. 1993). If one can
generalize these two examples, and if they provide snapshots in time of the
evolution of basically equal cores, it seems that a massive star forms first in
a relatively homogeneous core, another example with Plummer density profile and without fragmentation being the high-mass core G10.47  (Rolffs et al., in prep.), followed by fragmentation or at least visible
break-up of the core and subsequent star formation, perhaps aided by radiative
or outflow feedback from the first star. This scenario may well apply to extremely high-mass cluster forming cores or to special environments only, since high-mass IRDCs have been shown to fragment early (Zhang et al. 2009).

\end{document}